\documentclass[reprint,superscriptaddress,showpacs,nofootinbib,amsmath,amssymb]{revtex4-1}
\usepackage{comment}
\usepackage{amssymb}
\usepackage{color}
\usepackage{rotating}
\usepackage{amsmath}
\usepackage{ulem}
\usepackage{overpic}
\usepackage{graphicx}
\usepackage{dcolumn}
\usepackage{graphicx}
\usepackage{bm}
\usepackage{chngpage}
\usepackage{multirow}
\usepackage{slashed}
\usepackage{indentfirst}
\usepackage{booktabs}
\usepackage{amssymb,amsfonts}
\usepackage{amsmath}
\usepackage{color}
\usepackage[colorlinks,citecolor=blue,linkcolor=blue,hypertex,breaklinks=true]{hyperref}

\begin{document}

\title{Constraining tensor force terms with the charge radii difference of mirror-pair nuclei}

\author{Yan Ya}
\affiliation{School of Physics, Ningxia University, Yinchuan 750021, China}

\author{Na Tang}
\affiliation{School of Physics, Ningxia University, Yinchuan 750021, China}
\affiliation{Key Laboratory of Beam Technology of Ministry of Education, School of Physics and Astronomy, Beijing Normal University, Beijing 100875, China}

\author{Rong An}
\email[Contact author: ]{rongan@nxu.edu.cn}
\affiliation{School of Physics, Ningxia University, Yinchuan 750021, China}
\affiliation{Key Laboratory of Beam Technology of Ministry of Education, School of Physics and Astronomy, Beijing Normal University, Beijing 100875, China}
\affiliation{Guangxi Key Laboratory of Nuclear Physics and Technology, Guangxi Normal University, Guilin, 541004, China}

\begin{abstract}
 Charge radii differences of mirror partner nuclei provide an alternative probe to pin down the interaction components in asymmetric nuclear matter. In this work, the differences in the charge radii of almost spherical mirror-paired nuclei $^{54}$Ni-$^{54}$Fe and $^{36}$Ca-$^{36}$S are used to constrain the magnitude of tensor terms in the Skyrme interactions.
 The calculated results suggest that a linear correlation can be found between the difference of charge radii of mirror partner nuclei and the adopted strengths of the triplet-odd and triplet-even tensor components.
 Besides, it suggests that charge radii differences of mirror-paired nuclei are more sensitive to the adopted strengths of the triplet-odd parameter $U$ rather than the triplet-even parameter $T$.
 Combining the quantitative constraint strengths of the triplet-even tensor part obtained from the magnetic dipole (M1) excitations, the charge-exchange Gamow-Teller (GT) states, and the spin-dipole (SD) excitations, the triplet-odd strengths are further constrained for the SLy5 as well as SGII effective interactions.
 This provides an alternative approach to constrain the appropriate magnitude of tensor force.

\end{abstract}


\maketitle
\section{INTRODUCTION}\label{sec1}
The fundamental knowledge of nuclear force provides an access to understand the nature characteristic of strongly correlated many-body nucleons systems.
Particularly, as an important component of the bare nucleon-nucleon ($NN$) interaction, tensor force plays an indispensable role in nuclear physics and astrophysics.
Consequently, it can actually have an influence on determining the shell evolution through spin-orbit splitting of single particle states~\cite{PhysRevC.77.024316,PhysRevC.77.014314,PhysRevC.80.064302,PhysRevC.83.054305,Zhou_2012,BHATTACHARYA20131,PhysRevC.88.054328,PhysRevC.98.064318}.
The resultant nuclear phenomena, such as shell quenching effect~\cite{PhysRevC.89.034316,PhysRevC.89.064322,ANGUIANO2016181,PhysRevC.97.064304,PhysRevC.107.054307},
nuclear fission~\cite{PhysRevC.101.044615,PhysRevC.110.064318} or quasifission~\cite{LI2022137349,PhysRevC.110.064607}, heavy-ion collisions~\cite{PhysRevC.93.054617,PhysRevC.98.064607,PhysRevC.105.034601}, two-nucleon knockout reaction~\cite{PhysRevC.104.034306}, collective excited states~\cite{PhysRevC.80.064304,PhysRevC.92.044317,PhysRevLett.110.122501,PhysRevLett.116.089902,PhysRevC.103.054308}, two-proton radioactivity~\cite{PhysRevC.110.064305}, $\beta$-decay half-life~\cite{PhysRevC.107.014325,CPCadcf8f,PhysRevC.103.064327}, bubble structure~\cite{PhysRevC.84.044333,PhysRevC.87.067305,PhysRevC.91.017302,CPC114101}, and pseudospin symmetry (PSS)~\cite{PhysRevC.110.L051301}, etc, have been significantly influenced by the motivated tensor force interactions.
In addition, the tensor force contributing to the symmetry energy encourages us to investigate nuclear force from fundamental interactions as well~\cite{PhysRevC.84.062801,PhysRevC.91.025803,PhysRevC.111.064307}.

Tensor force, which contributes to the noncentral and nonlocal spin-orbit interaction, has been first discussed in the nonrelativistic Skyrme energy density functions (EDFs)~\cite{STANCU1977108}.
In the zero-range tensor interaction, the triplet-odd tensor parameter $U$ and the triplet-even one $T$ are actually used to characterize the effective interaction strengths.
This generally, as mentioned above, can be used to describe various physical phenomena well.
Nowadays, the role of tensor force component has been dramatically introduced into various models, such as shell model~\cite{PhysRevLett.95.232502}, ab initio approach~\cite{SHEN2018344}, self-consistent mean-field theory~\cite{PhysRevLett.97.162501,PhysRevC.76.014312,PhysRevC.101.064302}, etc.
Specifically, within the framework of relativistic mean field theory, the contribution of the tensor force is naturally induced from the Fock diagrams~\cite{PhysRevC.91.025802,PhysRevC.103.064326,PhysRevC.105.034329}.
As demonstrated in Ref.~\cite{PhysRevC.98.034313}, the quantitative analysis of tensor effects in the relativistic Hartree-Fock (RHF) theory has been performed.
It is found that all the meson-nucleon couplings, except the $\sigma$-scalar one, give rise to the tensor force counterparts.
Particularly, the impact of the tensor force in neutron-proton drops and the evolution of single-particle energies along the Ca and Sn isotopic chains have been investigated using the RHF theory with considering $\pi$-exchange potential~\cite{LOPEZQUELLE2018149,PhysRevC.100.064319,PhysRevC.102.034322}.

Recently, tensor correlation strength has brought more attention in nuclear structure phenomena.
The interaction strength of tensor parameters can be fixed by reproducing the single-particle energy differences due to the effect of tensor on the spin-orbit splitting~\cite{PhysRevC.74.061303,PhysRevC.75.064311,COLO2007227,PhysRevC.89.044314,PhysRevC.95.034307}.
Besides, some collective excited states, such as spin-isospin excitations~\cite{PhysRevC.81.044302}, Gamow-Teller (GT) states~\cite{BAI200928}, charge-exchange spin-dipole (SD) mode~\cite{PhysRevLett.105.072501}, magnetic dipole modes~\cite{PhysRevC.83.034324,PhysRevC.89.044311,PhysRevC.98.064312,PhysRevC.109.014321}, are sensitive to the tensor components in two-body particle-particle interaction channels.
As mentioned in Ref.~\cite{PhysRevC.83.054316}, the GT and SD excitation energies in $^{90}$Zr and $^{208}$Pb are systematically studied to constrain the appropriate magnitude of the tensor terms on top of the existing effective Skyrme interactions.
Tensor effect in shell evolution has also been addressed by Moreno-Torres et al.~\cite{PhysRevC.81.064327} where
they have suggested that the magic numbers $8$ and $20$ are suitable for fitting the tensor parameters in mean-field approach.
The optimized tensor-force strength $\lambda$ in the RHF theory, which reproduces the relativistic Brueckner-Hartree-Fock (RBHF) spin-orbit splitting, is running with the strength of the external fields of neutron drops. This provides an important guide for future determination of tensor forces in nuclear EDFs based on microscopic $ab~initio$ calculations~\cite{PhysRevC.100.064319}.
Recent study suggests that the triplet-even tensor strength $T$ can be constrained well, but the triplet-odd one $U$ still cannot be determined~\cite{99xk-tm1q}.
This means that the role of appropriate tensor strength is still an open issue and is fairly uncertain.
Therefore, more potential features should be required in capturing the information about tensor strength of the existing parameters used in Skyrme EDFs.

As mentioned before, tensor force can actually have an influence on determining the evolution of single-particle levels, namely affecting the density distributions of neutron and proton.
Recently, a highly linear correlation between the charge radii difference ($\Delta{\mathrm{R_{ch}}}$) of mirror partner nuclei and the slope parameter ($L$) of symmetry energy has been undertaken to determine the isospin interaction components in the equation of state (EoS) of asymmetric nuclear matter~\cite{PhysRevC.88.011301,PhysRevLett.119.122502,PhysRevC.97.014314,PhysRevResearch.2.022035,XU2022137333,PhysRevLett.127.182503,PhysRevLett.130.032501,nuclscitech34.119,
PhysRevC.108.015802,PhysRevLett.132.162502}.
The mass-dependent linear relationship between $\Delta{\mathrm{R_{ch}}}$ of mirror pairs and isospin asymmetry degrees is naturally emerged~\cite{PhysRevC.110.014316}.
Inspired by these motivations, it is instructive to investigate the correlation between the tensor force and the difference of charge radii of mirror-pair nuclei.
In this work, the spherical mirror partner nuclei $^{54}$Ni-$^{54}$Fe and $^{36}$Ca-$^{36}$S have been used to investigate the quantified range of the tensor force strength parameters with the framework of Skyrme EDFs.
The effective forces SGII~\cite{VANGIAI1981379} and SLy5~\cite{CHABANAT1998231} are employed in our discussion.

The structure of the paper is organized as follows.
In Sec.~\ref{sec2}, we succinctly present the theoretical framework.
In Sec.~\ref{sec3}, the correlations between the charge radii difference of mirror partner nuclei $^{54}$Ni-$^{54}$Fe and $^{36}$Ca-$^{36}$S and the triplet-even tensor strength parameter $T$ and triplet-odd tensor strength parameter $U$ are also discussed with Skyrme EDFs using the mixed-pairing interaction forces, respectively.
Meanwhile, combining the existing results in the literature, the triplet-odd strengths $U$ are further constrained for the SLy5 as well as SGII effective interactions.
Finally, a summary and outlook are given in Sec.~\ref{sec4}.

\section{Theoretical Framework}\label{sec2}
Many properties of unstable nuclei are described successfully within the Skyrme-Hartree-Fock-Bogoliubov (SHFB) theory including the tensor force~\cite{BONCHE200549,BENNACEUR200596,DOBACZEWSKI20092361,STOITSOV20131592}.
In this work, the Skyrme-like effective interaction has been recalled as follows~\cite{CHABANAT1997710,CHABANAT1998231},
\begin{eqnarray}
V(\mathbf{r}_{1},\mathbf{r}_{2})&=&t_{0}(1+x_{0}\mathbf{P}_{\sigma})\delta(\mathbf{r})\nonumber\\
&&+\frac{1}{2}t_{1}(1+x_{1}\mathbf{P}_{\sigma})\left[\mathbf{P}'^{2}\delta(\mathbf{r})+\delta(\mathbf{r})\mathbf{P}^{2}\right]\nonumber\\
&&+t_{2}(1+x_{2}\mathbf{P}_{\sigma})\mathbf{P}'\cdot\delta(\mathbf{r})\mathbf{P}\nonumber\\
&&+\frac{1}{6}t_{3}(1+x_{3}\mathbf{P}_{\sigma})[\rho(\mathbf{R})]^{\alpha}\delta(\mathbf{r})\nonumber\\
&&+\mathrm{i}W_{0}\mathbf{\sigma}\cdot\left[\mathbf{P}'\times\delta(\mathbf{r})\mathbf{P}\right].
\end{eqnarray}
Here, $\mathbf{r}=\mathbf{r}_{1}-\mathbf{r}_{2}$ and $\mathbf{R}=(\mathbf{r}_{1}+\mathbf{r}_{2})/2$ are related to the positions of two nucleons $\mathbf{r}_{1}$ and $\mathbf{r}_{2}$, $\mathbf{P}=(\nabla_{1}-\nabla_{2})/2\mathrm{i}$ is the relative momentum operator and $\mathbf{P'}$ is its complex conjugate acting on the left, and $\mathbf{P_{\sigma}}=(1+\vec{\sigma}_{1}\cdot\vec{\sigma}_{2})/2$ is the spin exchange operator that controls the relative strength of the $S=0$ and $S=1$ channels for a given term in the two-body
interactions, with $\vec{\sigma}_{1(2)}$ being the Pauli matrices.
The last term represents the spin-orbit force where $\sigma=\vec{\sigma}_{1}+\vec{\sigma}_{2}$.
The quantities $\alpha$, $t_{0}$, $t_{1}$, $t_{2}$, $t_{3}$, $x_{0}$, $x_{1}$, $x_{2}$, and $x_{3}$ represent the parameters of the Skyrme forces used in this work, $W_{0}$ being the spin-orbit strength parameter.

The triplet-even and triplet-odd zero-range tensor force, which originated from the form given by T. H. R Skyrme and read~\cite{STANCU1977108}
\begin{eqnarray}
V^{T}&=&\frac{T}{2}\bigg\{\left[(\mathbf{\sigma_{1}}\cdot{\mathbf{P'}})(\mathbf{\sigma_{2}}\cdot{\mathbf{P'}})
-\frac{1}{3}(\sigma_{1}\cdot\sigma_{2})\mathbf{P'^{2}}\right]\delta(\mathbf{r_{1}}-\mathbf{r_{2}})\nonumber\\
&&+\delta(\mathbf{r_{1}}-\mathbf{r_{2}})\left[(\mathbf{\sigma_{1}}\cdot{\mathbf{P}})(\mathbf{\sigma_{2}}\cdot{\mathbf{P}})
-\frac{1}{3}(\sigma_{1}\cdot\sigma_{2})\mathbf{P^{2}}\right]\bigg\}\nonumber\\
&&+\frac{U}{2}\bigg\{(\mathbf{\sigma_{1}}\cdot{\mathbf{P'}})\delta(\mathbf{r_{1}}-\mathbf{r_{2}})(\mathbf{\sigma_{2}}\cdot{\mathbf{P}})\nonumber\\
&&+(\mathbf{\sigma_{2}}\cdot{\mathbf{P'}})\delta(\mathbf{r_{1}}-\mathbf{r_{2}})\times(\mathbf{\sigma_{1}}\cdot{\mathbf{P}})\nonumber\\
&&-\frac{2}{3}[(\mathbf{\sigma_{1}}\cdot\mathbf{\sigma_{2}})\mathbf{P'}\cdot
\delta(\mathbf{r_{1}}-\mathbf{r_{2}})\mathbf{P}]\bigg\}.
\end{eqnarray}
The coupling constants $T$ and $U$ denote the strengths of the triplet-even and triplet-odd tensor interactions, respectively. The central exchange and tensor contributions to the energy density $H(\mathbf{r})$ are
\begin{eqnarray}
\Delta{H}=\frac{1}{2}\alpha(\mathbf{J}_{n}^{2}+\mathbf{J}_{p}^{2})+\beta{\mathbf{J}_{n}\mathbf{J}_{p}},
\end{eqnarray}
where $\mathbf{J}_{n}$ and $\mathbf{J}_{p}$ are spin-orbit densities for neutrons and protons, respectively, defined by
\begin{small}
\begin{eqnarray}
\mathbf{J}_{q}(\mathbf{r})=\frac{1}{4\pi{r^{3}}}\sum_{i}(2j_{i}+1)\left[j_{i}(j_{i}+1)-l_{i}(l_{i}+1)-\frac{3}{4}\right]R_{i}^{2}(\mathbf{r}).\nonumber\\
\end{eqnarray}
\end{small}
In the above equation, the isospin quantum number $q=n(p)$ labels neutrons(protons), whereas the index $i=n, l, j$ runs over all occupied states having the given $q$, and $R_{i}(\mathbf{r})=u_{i}(\mathbf{r})/r$ is the radial part of the wavefunction. The spin-orbit potential is given by
\begin{eqnarray}
U_{\mathrm{SO}}^{(q)}=\frac{W_{0}}{2r}\left(2\frac{d\rho_{q}}{d{r}}+\frac{d\rho_{q'}}{dr}\right)
+\left(\alpha\frac{J_{q}}{r}+\beta\frac{J_{q'}}{r}\right),
\end{eqnarray}
where the first term comes from the Skyrme two-body spin-orbit interaction, whereas the second term includes both the central exchange and the tensor contributions, that is
\begin{eqnarray}\label{alpha}
\alpha=\alpha_{C}+\alpha_{T},~~~~\beta=\beta_{C}+\beta_{T}
\end{eqnarray}
with
\begin{eqnarray}
\alpha_{C}&=&\frac{1}{8}(t_{1}-t_{2})-\frac{1}{8}(t_{1}x_{1}+t_{2}x_{2}),\nonumber\\
\beta_{C}&=&-\frac{1}{8}(t_{1}x_{1}+t_{2}x_{2}),\nonumber\\
\alpha_{T}&=&\frac{5}{12}U,\nonumber\\
\beta_{T}&=&\frac{5}{24}(T+U).\nonumber
\end{eqnarray}
It should be noted that $J_{q}$ gives essentially no contribution in the spin-orbit saturated ($\mathbf{l}\cdot{\mathbf{s}}$ closed) nuclei in which the two spin-orbit partners are both occupied or unoccupied. Therefore, the tensor force gives a negligible contribution to the energy density in nuclei that are spin-orbit saturated for both neutrons and protons.

The charge radius is calculated using the following formula:
\begin{eqnarray}\label{cp1}
R_{\mathrm{ch}}^{2}=\langle{r_{\mathrm{p}}^{2}}\rangle+0.7056.
\end{eqnarray}
The first term represents the charge distribution of point-like protons and the second term accounts for the finite size effects of the proton. Here, the quantity of the proton radius takes the value about $0.84$ fm~\cite{RevModPhys.93.025010,PhysRevLett.128.052002}.

As is well known, pairing correlation plays an important role in describing the bulk properties of finite nuclei, especially the charge radii difference of mirror-paired nuclei~\cite{PhysRevC.105.L021301,PhysRevC.107.034319,GAUTAM2024122832}. In our calculations, the zero-range delta force is used to tackle pairing correlations around the Fermi surface. It should also be mentioned that zero-range pairing forces can be density dependent as follows~\cite{BERTSCH1991327,PhysRevC.71.054303}:
\begin{eqnarray}
V(\mathbf{r}-\mathbf{r'})=V_{0}\left\{1-\eta\left[\frac{\rho(\mathbf{r})}{\rho_{0}}\right]^{\alpha}\right\}\delta(\mathbf{r}-\mathbf{r'}).
\end{eqnarray}
Here, $V_{0}$ represents the pairing strength and the quantity $\rho_{0}=0.16$ denotes the saturation density. Generally, the values of $\eta$ are taken as $0.0$, $0.5$, and $1.0$ for volume-, mixed-, and surface-type pairing interactions, respectively. Furthermore, $\rho(\mathbf{r})$ is the baryon density distribution in coordinate space and $\alpha$ is taken as $1.0$ in our discussion.
The pairing window of $30.0$ MeV both above and below the Fermi surface is employed for the particle-particle channel.
The quantity $V_{0}$ is generally adjusted by calibrating the empirical binding energy gaps with three-point formula~\cite{Ring1980,JANECKE200323}. In order to reflect the universality of our results, the mixed-type pairing interactions are used in our discussion.

\section{Results and discussion}\label{sec3}
As mentioned in Refs.~\cite{PhysRevC.88.011301,PhysRevLett.119.122502,PhysRevC.97.014314,PhysRevResearch.2.022035,XU2022137333,PhysRevLett.127.182503,PhysRevLett.130.032501,nuclscitech34.119,
PhysRevC.108.015802,PhysRevLett.132.162502}, a highly linear correlation can be found between the charge radii difference ($\Delta{R}_{\mathrm{ch}}$) of mirror-pair nuclei and the slope parameter of symmetric nuclear matter.
As the aforementioned discussion, tensor force can actually affect the single-particle energies due to the spin-orbit splitting interactions~\cite{PhysRevC.89.034316,PhysRevC.89.064322,PhysRevC.97.064304,PhysRevC.107.054307}.
This further leads to the changes of spatial density distributions for neutron and proton components, namely the changes in neutron and proton matter radii.
Inspired by these motivations, the correlation between the charge radii difference of mirror nuclei and tensor strength parameters should be investigated.
Meanwhile, the influences coming from shape deformation effects~\cite{CPC094101} and isoscalar compression modulus~\cite{nuclscitech35.182,Ma_2024} can be encountered in discussing the difference of charge radii of mirror nuclei.
To avoid the influence coming from the deformation effect, the almost spherical mirror-paired nuclei $^{54}$Ni-$^{54}$Fe and $^{36}$Ca-$^{36}$S are chosen. The tensor interactions are obtained by adding tensor terms on top of the existing effective forces, SGII~\cite{VANGIAI1981379} and SLy5~\cite{CHABANAT1998231}.

\begin{figure}[htbp]
\includegraphics[scale=0.33]{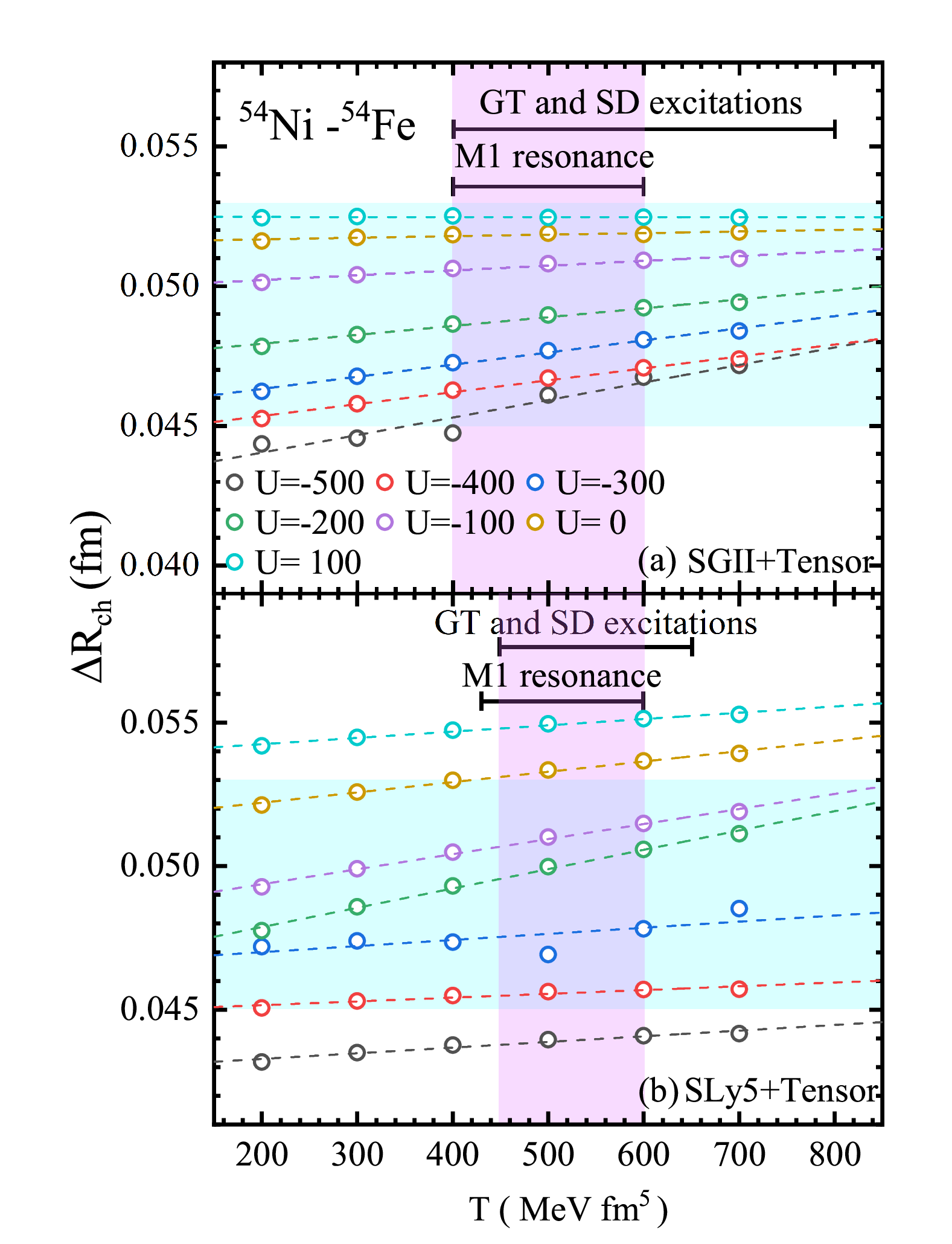}
\caption{$\Delta{R}_{\mathrm{ch}}$ of mirror nuclei $^{54}$Ni-$^{54}$Fe as a function of the triplet-even tensor strength $T$ is depicted by the effective forces SGII~\cite{VANGIAI1981379} (a) and SLy5~\cite{CHABANAT1998231} (b) with the mixed-type pairing force, respectively. The experimental result $\Delta{R}_{\mathrm{ch}}$ is shown as a horizontal light blue band~\cite{PhysRevLett.127.182503}. The quantitatively constrained range of the triplet-even tensor strengths $T$ obtained from the charge-exchange Gamow-Teller (GT) and the spin-dipole (SD) excitations~\cite{PhysRevC.83.054316} and the magnetic dipole (M1) excitations~\cite{99xk-tm1q} are also labeled by light purple band. The dashed lines indicate theoretical linear fits with the fixed triplet-odd tensor strength $U$.} \label{fig1}
\end{figure}
As shown in Fig.~\ref{fig1}, the correlations between the $\Delta{R}_{\mathrm{ch}}$ of mirror nuclei $^{54}$Ni-$^{54}$Fe and the triplet-even tensor strength $T$ classified by various triplet-odd tensor strength $U$ are clearly presented by the effective forces SGII~\cite{VANGIAI1981379} and SLy5~\cite{CHABANAT1998231} with the mixed-type pairing force, respectively.
The experimental value, $\Delta{R}_{\mathrm{ch}}$ =0.049(4) fm for mirror nuclei $^{54}$Ni-$^{54}$Fe~\cite{PhysRevLett.127.182503}, is also displayed for comparison in Figs.~\ref{fig1}~(a) and (b).
For the fixed triplet-odd tensor strength $U=100$ MeV fm$^{5}$ and $U=0$ MeV fm$^{5}$, as shown in Fig.~\ref{fig1}~(a), the theoretical linear fits are almost parallel to the x-axis. This means that the triplet-even tensor strength $T$ is not sensitive to the $\Delta{R}_{\mathrm{ch}}$ of mirror nuclei $^{54}$Ni-$^{54}$Fe. From the fixed triplet-odd tensor strength parameters $U=-500$ MeV fm$^{5}$ to $U=-100$ MeV fm$^{5}$, a slightly increasing trend in $\Delta{R}_{\mathrm{ch}}$ can be observed with the increasing triplet-even tensor strength $T$.
The same scenario can also be encountered for the SLy5 force as shown in Fig.~\ref{fig1}~(b).

\begin{figure}[htbp]
\includegraphics[scale=0.33]{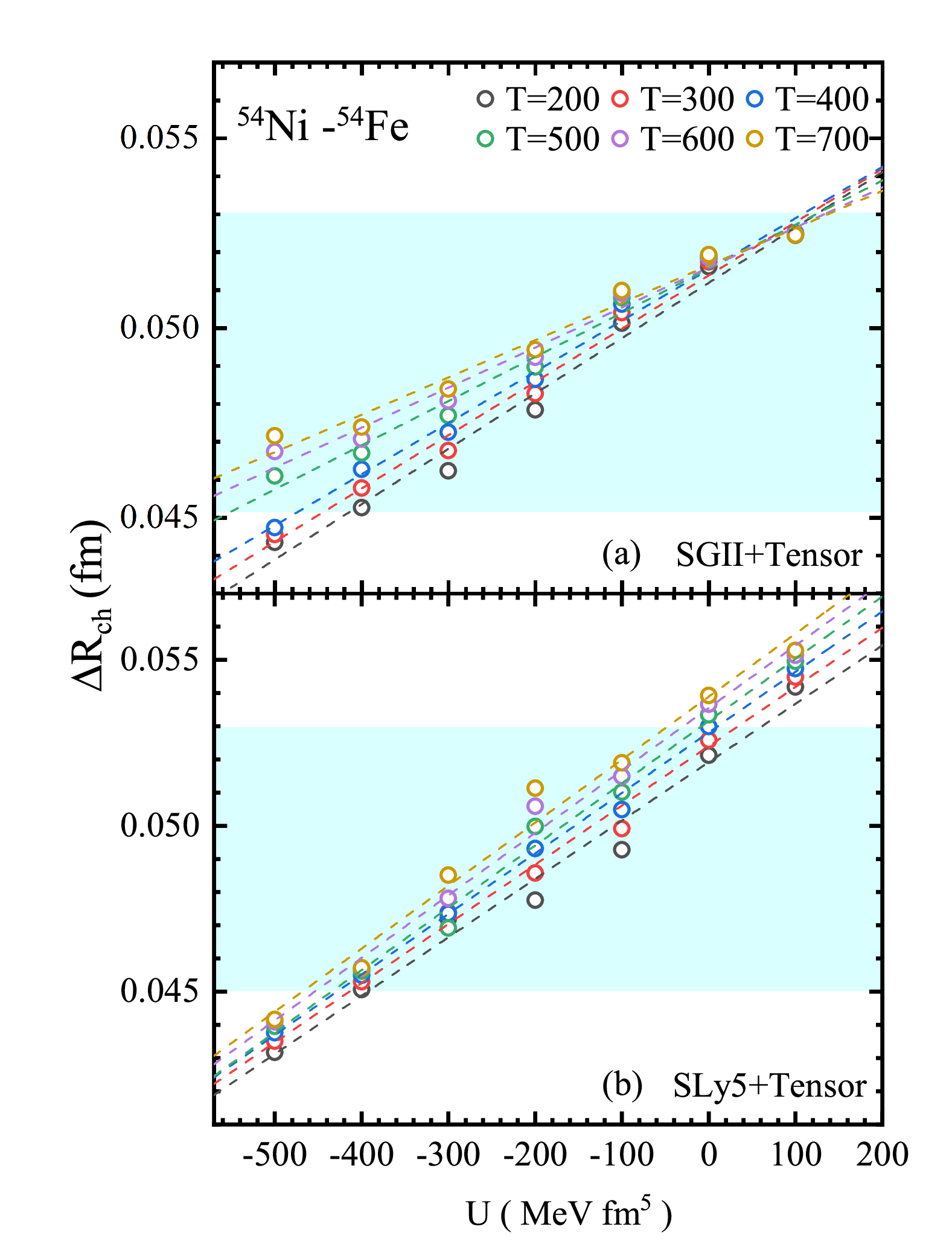}
\caption{$\Delta{R}_{\mathrm{ch}}$ of mirror nuclei $^{54}$Ni-$^{54}$Fe as a function of the triplet-odd tensor strength $U$ is depicted by the effective forces SGII~\cite{VANGIAI1981379} (a) and SLy5~\cite{CHABANAT1998231} (b) with the mixed-type pairing force, respectively. The experimental value of $\Delta{R}_{\mathrm{ch}}$ is shown as a horizontal light blue band~\cite{PhysRevLett.127.182503}. The dashed lines indicate theoretical linear fits with the fixed triplet-even tensor strength $T$.} \label{fig2}
\end{figure}
From this figure, it should be remarkably mentioned that linear correlations can be significantly observed between the values of $\Delta{R}_{\mathrm{ch}}$ of mirror nuclei $^{54}$Ni-$^{54}$Fe and the triplet-even tensor strength $T$ classified by various fixed triplet-odd tensor strength parameters $U$.
These calibrated linear fits can cover the experimental uncertainty $\Delta{R}_{\mathrm{ch}}$ of mirror nuclei $^{54}$Ni-$^{54}$Fe.
However, the slope of the fitting lines is rather small, even some slopes of the theoretical linear fits tend to zero.
This leads to the wide range of the triplet-even tensor strength $T$ with respect to the quantitatively constrained triplet-even tensor strengths $T$ derived from the charge-exchange Gamow-Teller (GT) and the spin-dipole (SD) excitations~\cite{PhysRevC.83.054316} and the magnetic dipole (M1) excitations~\cite{99xk-tm1q}.
This means that the triplet-even tensor strength $T$ cannot be constrained by the $\Delta{R}_{\mathrm{ch}}$ of mirror nuclei $^{54}$Ni-$^{54}$Fe well.

\begin{figure}[htbp]
\includegraphics[scale=0.33]{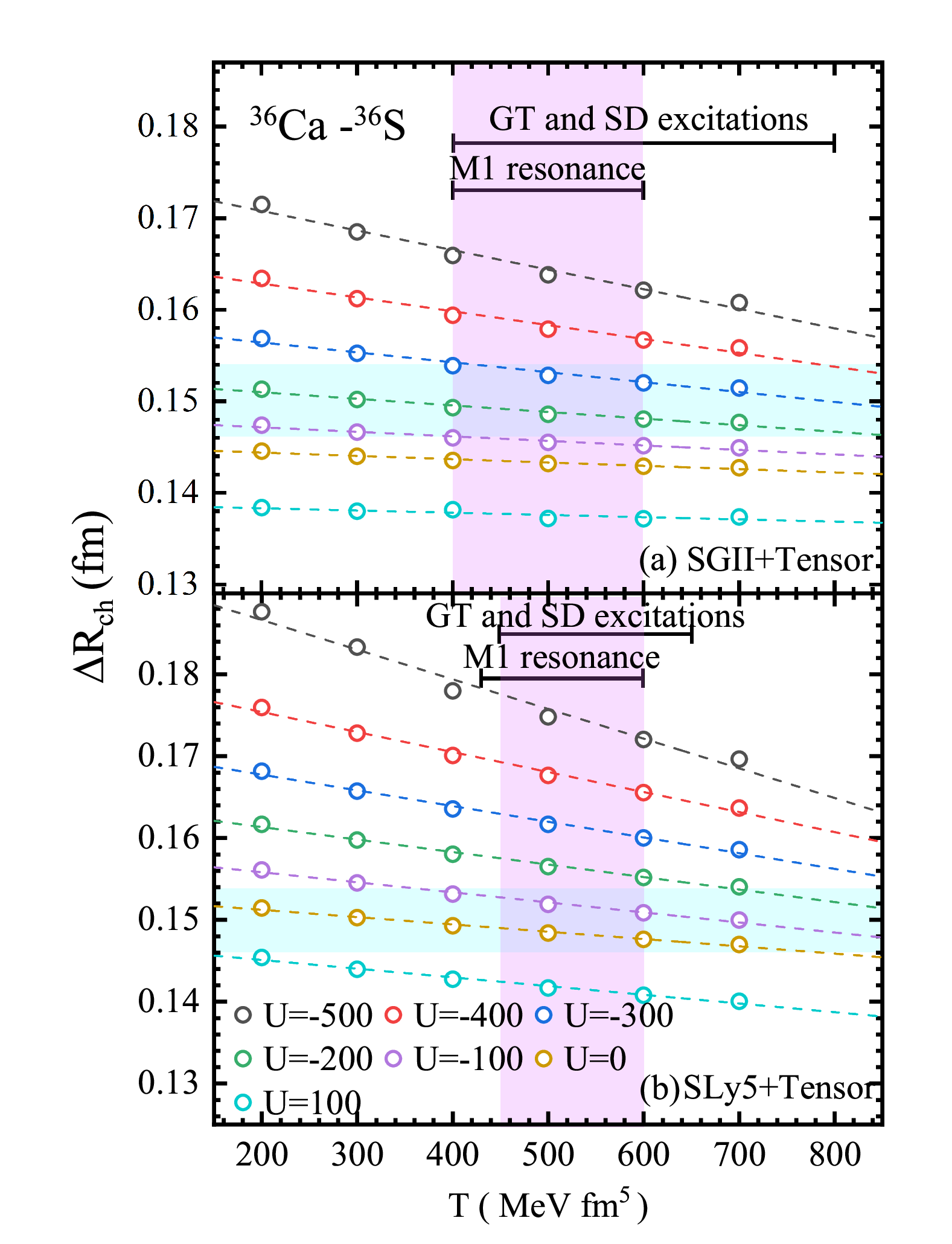}
\caption{Same as Fig.~\ref{fig1}, but for mirror-paired nuclei $^{36}$Ca-$^{36}$S.} \label{fig3}
\end{figure}
In Fig.~\ref{fig2}, $\Delta{R}_{\mathrm{ch}}$ of mirror nuclei $^{54}$Ni-$^{54}$Fe as a function of the triplet-odd tensor strength $U$ is also depicted by the SGII and SLy5 forces with various fixed triplet-even tensor strength $T$.
It should be mentioned that linear correlations between the $\Delta{R}_{\mathrm{ch}}$ of mirror nuclei $^{54}$Ni-$^{54}$Fe and the triplet-even tensor parameter $U$ classified by the specific $T$ values are clearly presented.
Here, it should be noted that the slope of the theoretically fitting line is rather larger than that encountered in the triplet-even tensor strength $T$ as shown in Fig.~\ref{fig1}.
This further indicates that the triplet-odd tensor strength $U$ is more sensitive to the $\Delta{R}_{\mathrm{ch}}$ of mirror nuclei $^{54}$Ni-$^{54}$Fe rather than the triplet-even tensor counterparts.
This seems to provide an alternative access to constrain the triplet-odd tensor strength $U$.

\begin{figure}[htbp]
\includegraphics[scale=0.33]{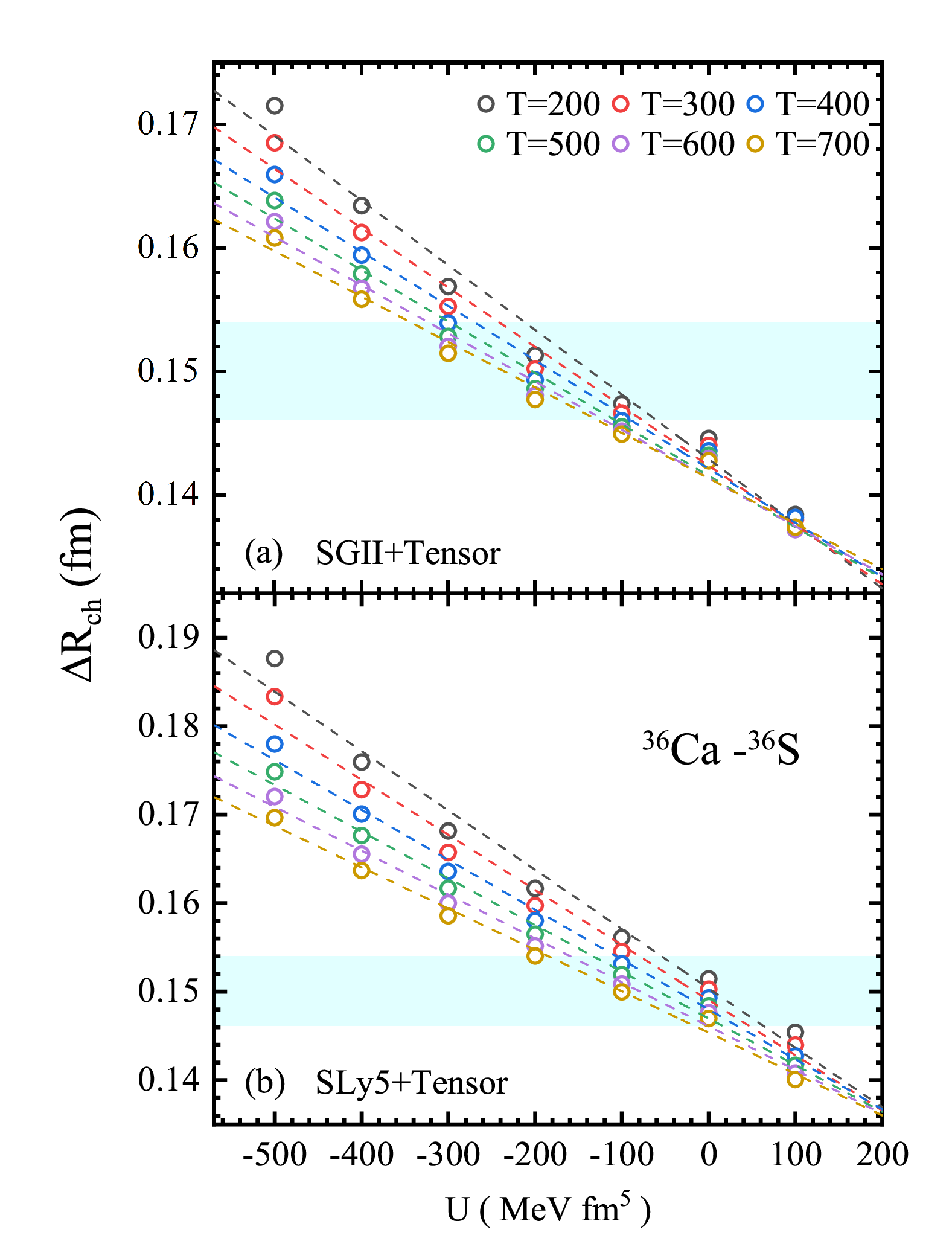}
\caption{Same as Fig.~\ref{fig2}, but for mirror-paired nuclei $^{36}$Ca-$^{36}$S.} \label{fig4}
\end{figure}
To further review this phenomenon, as shown in Fig.~\ref{fig3}, the correlations between the $\Delta{R}_{\mathrm{ch}}$ of mirror nuclei $^{36}$Ca-$^{36}$S and the triplet-even tensor strength $T$ are also displayed by the effective forces SGII and SLy5 with the mixed-type pairing force, respectively. The experimental $\Delta{R}_{\mathrm{ch}}$ of mirror nuclei $^{36}$Ca-$^{36}$S is 0.150$\pm$0.004 fm~\cite{PhysRevResearch.2.022035}. As encountered in Fig.~\ref{fig1}, the linear correlation can also be presented, but the trend is decreased with the increasing triplet-even tensor parameter $T$.
Meanwhile, it should be mentioned that the slopes of the fitting lines are gradually close to zero from $U=-500$ MeV fm$^{5}$ to 100 MeV fm$^{5}$ as shown in Fig.~\ref{fig3}~(a). In Fig.~\ref{fig3}~(b), the changes of the slopes in the theoretical fitting lines are rather small except the case encountered in $U=-500$ MeV fm$^{5}$ and $U=-400$ MeV fm$^{5}$.
Compared to the results obtained from the charge-exchange Gamow-Teller (GT) and the spin-dipole (SD) excitations~\cite{PhysRevC.83.054316} and the magnetic dipole (M1) excitations~\cite{99xk-tm1q}, the covered range of the triplet-even tensor components $T$ cannot be determined well through the value of $\Delta{R}_{\mathrm{ch}}$ of mirror nuclei $^{36}$Ca-$^{36}$S.
This is in accord with the results shown in Fig.~\ref{fig1}.

Besides, as seen in Fig.~\ref{fig4}, $\Delta{R}_{\mathrm{ch}}$ of mirror nuclei $^{36}$Ca-$^{36}$S as a function of the triplet-odd tensor strength $U$ classified by various fixed triplet-even tensor strength $T$ is also depicted with the SGII and SLy5 forces, respectively.
From this figure, it can be noted that the reverse linear relationship can be obtained between the values of $\Delta{R}_{\mathrm{ch}}$ of mirror nuclei $^{36}$Ca-$^{36}$S and the triplet-odd tensor strength parameter $U$.
Compared to the relations between the $\Delta{R}_{\mathrm{ch}}$ of mirror nuclei $^{36}$Ca-$^{36}$S and the triplet-even tensor strength $T$ shown in Fig.~\ref{fig3}, it is found that the triplet-odd tensor strength parameter $U$ is more sensitive to the values of $\Delta{R}_{\mathrm{ch}}$ of mirror nuclei $^{36}$Ca-$^{36}$S.
This is in accord with that encountered in mirror nuclei $^{54}$Ni-$^{54}$Fe.
As suggested in Ref.~\cite{PhysRevC.76.014312}, the so-called $TIJ$ ($I, J=1, \cdots, 6$) family forces which include the central and tensor terms on equal footing were proposed by a variational procedure. Here, it should be mentioned that the parametrization components in the effective forces $TIJ$, such as $t_{0}$, $t_{1}$, et al., are changed. Actually, linear correlations can also be obtained between $\Delta{R}_{\mathrm{ch}}$ of mirror-paired nuclei and the triplet-odd tensor strengths $U$ deduced from the $TIJ$ family forces.

All calculations suggest that a linear correlation can be observed between the charge radii differences of mirror nuclei and the triplet-even tensor strength $T$ classified by various fixed triplet-odd tensor strength parameter $U$. This linear correlation can also be characterized between the $\Delta{R}_{\mathrm{ch}}$ of mirror nuclei and the triplet-odd tensor strength $U$ classified by various fixed triplet-even tensor strength parameter $T$. As demonstrated in Refs.~\cite{PhysRevC.83.054316,99xk-tm1q}, the triplet-even tensor strength $T$ depends on the triplet-odd one $U$. From figures~\ref{fig1} and~\ref{fig3}, one can note that the covered range of $T$ is actually influenced by the values of $U$. However, compared these results shown in Figs.~\ref{fig1}-\ref{fig4}, it suggests that the triplet-odd tensor strength $U$ is more sensitive to the charge radii differences of mirror nuclei rather than the triplet-even tensor strength parameter $T$. This means that the triplet-odd tensor strength $U$ can be actually determined by using the values of $\Delta{R}_{\mathrm{ch}}$ of mirror nuclei.

Meanwhile, a positive correlation between $\Delta{R_{ch}}$ and $U$ for $^{54}$Ni-$^{54}$Fe is obtained as shown in Fig.~2, while the opposite correlation shown in Fig.~4 is predicted for $^{36}$Ca-$^{36}$S. As is well known, tensor force has an influence on determining the shell evolution~\cite{PhysRevC.77.024316,PhysRevC.77.014314,PhysRevC.80.064302,PhysRevC.83.054305,Zhou_2012,BHATTACHARYA20131,PhysRevC.88.054328,PhysRevC.98.064318}, namely influencing the shell gaps. In our calculations, the $sd$-shell for mirror-paired nuclei $^{36}$Ca-$^{36}$S and $pf$-shell for $^{54}$Ni-$^{54}$Fe are dramatically influenced by adding the  tensor forces. With the increasing triplet-odd tensor strength $U$, more particles are partially scattered into the $1d_{3/2}$ shell for $^{36}$S. This leads to the enlarged charge radius. However, it should be mentioned that the $1d_{3/2}$ and $2s_{1/2}$ shells are fully occupied for $^{36}$Ca, namely the charge radius is almost unchanged. Therefore, the values of $\Delta{R_{ch}}$ of mirror-paired nuclei $^{36}$Ca-$^{36}$S are gradually decreased with the increasing $U$. For $^{54}$Ni-$^{54}$Fe, the protons are gradually scattered into the $2p_{3/2}$ and $2p_{1/2}$ shells with the increasing triplet-odd tensor strength $U$. However, compared to $^{54}$Fe, more protons are scattered into the $p$-shell for $^{54}$Ni. This means that $\Delta{R}_{ch}$ of mirror-paired nuclei $^{54}$Ni-$^{54}$Fe is gradually increased with the increasing triplet-odd tensor strength $U$, namely it leads to positive correlation.

As mentioned in Ref.~\cite{PhysRevC.83.054316}, the triplet-even tensor strength $T$ covers the range from $450$ MeV fm$^{5}$ to $650$ MeV fm$^{5}$ based on the framework of force SLy5. Meanwhile, the triplet-odd tensor strength $U$ ranges from $-300$ MeV fm$^{5}$ to $-50$ MeV fm$^{5}$. Actually, the maximum value of triplet-even tensor strength $T$ depends on the triplet-odd counterpart. For SGII interaction force, the value of $T$ is limited within an area specified by $400<T<800$ MeV fm$^{5}$. Here, it should be noted that the triplet-even tensor strength $T$ depends on the value of $U$, and the quantity $U$ covers the negative range from $-280$ MeV fm$^{5}$ to $-80$ MeV fm$^{5}$ under the effective SGII interaction.
\begin{figure}[htbp]
\includegraphics[scale=0.3]{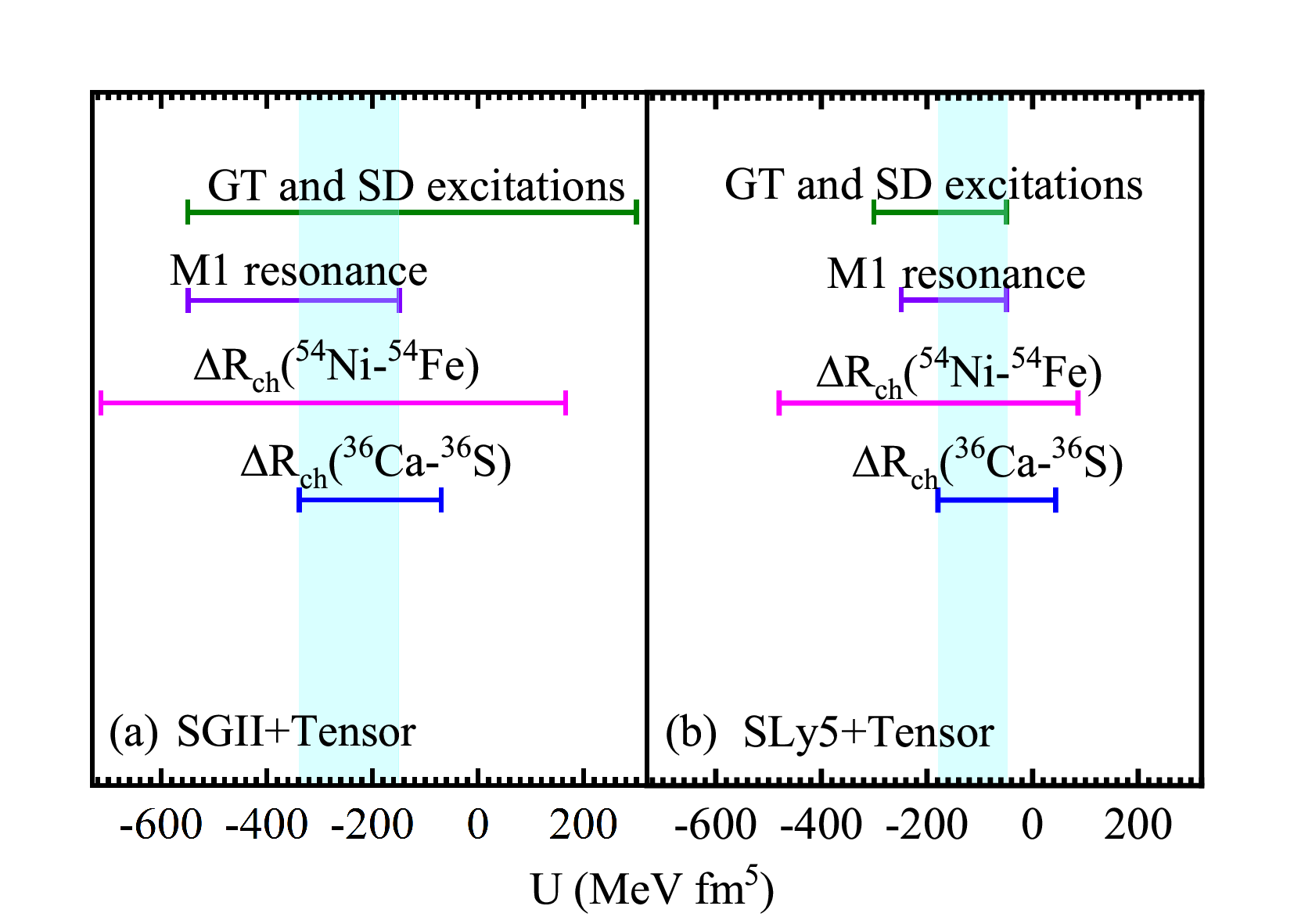}
\caption{Comparison between the values of the triplet-odd strength parameter $U$ extracted from the charge radii differences of mirror nuclei $^{54}$Ni-$^{54}$Fe and $^{36}$Ca-$^{36}$S and those obtained from the charge-exchange Gamow-Teller (GT) and the spin-dipole (SD) excitations~\cite{PhysRevC.83.054316} and the magnetic dipole (M1) excitations~\cite{99xk-tm1q}. The light blue regions are labeled as the further constrained range with various methods.} \label{fig5}
\end{figure}
The M1 resonance of $^{124}$Sn further reduces the upper limit of $T$ from $800$ MeV fm$^{5}$ to $600$ MeV fm$^{5}$ for SGII interaction force~\cite{99xk-tm1q}. For the effective SLy5 force, the lower limit of $T$ is limited about $400$ MeV fm$^{5}$ and the maximum value of $T$ is restricted to $600$ MeV fm$^{5}$.
For SGII interaction force, using the $M1$ excitation energy of $^{124}$Sn, which can achieve a quite effective restriction on the values of the parameter $U$ and further narrow its range from $(-550, 300)$ MeV fm$^{5}$ down to $(-550,-150)$ MeV fm$^{5}$.
For effective SLy5 force, the parameter $U$ can vary from $-300$ MeV fm$^{5}$ to $-50$ MeV fm$^{5}$. Meanwhile, the upper limit of the  triplet-even tensor strength $T$ depends on $U$, especially the maximum value of $T$ is close to $600$ MeV fm$^{5}$ when $U$ equals to $-190$ MeV fm$^{5}$.

As encountered in Refs.~\cite{PhysRevC.83.054316,99xk-tm1q}, the covered range of $T$ depends on $U$. In our calculations, the charge radii differences of mirror partner nuclei are more sensitive to the triplet-odd tensor strength parameter $U$ rather than the triplet-even components. As shown in Figs.~\ref{fig1} and~\ref{fig3}, the covered range of $T$ is relatively large. To determine the appropriate magnitude tensor strength in our simulations, the validated ranges of $T$ are limited at the $(450, 600)$ MeV fm$^{5}$ and $(400, 600)$ MeV fm$^{5}$ for the corresponding SLy5 and SGII forces, respectively.
As shown in Figs.~\ref{fig2} and~\ref{fig4}, linear functions can be fitted to the triplet-odd tensor strengths $U$ under the fixed $T$ values. The constrained range of the triplet-odd tensor strength $U$ is derived from the intersecting regions as shown in Figs.~\ref{fig2} and~\ref{fig4} where the experimental $\Delta{R}_{ch}$ bands overlap the theoretically fitting lines with the specific $T$ ranges.
As shown in Fig.~\ref{fig5}, the comparison results are presented with these methods.
For SGII force, the covered ranges of $U$ derived from mirror-pair nuclei $^{54}$Ni-$^{54}$Fe and $^{36}$Ca-$^{36}$S are $(-714, 165)$ MeV fm$^{5}$ and $(-341, -73)$ MeV fm$^{5}$, respectively. In contrast, the corresponding triplet-odd tensor strengths are limited at the range of $(-480, 85)$ MeV fm$^{5}$ and $(-182, 42)$ MeV fm$^{5}$ for the mirror-pair nuclei $^{54}$Ni-$^{54}$Fe and $^{36}$Ca-$^{36}$S in the effective SLy5 set.
Constraints on $U$ are deduced by comparing the theoretically fitting lines with the experimental $\Delta{R_{ch}}$ results of mirror-paired nuclei. Actually, the constrained tensor force can be used to simultaneously describe well the $\Delta{R_{ch}}$ of mirror-paired nuclei in different shell regions, although the inverse linear correlations have been predicted as shown in Figs.~\ref{fig2} and~\ref{fig4}.

Combining the validated ranges of $U$ obtained from the magnetic dipole (M1) excitations, the charge-exchange Gamow-Teller (GT) states, and the spin-dipole (SD) excitations, the triplet-odd tensor strength $U$ can be further constrained in this work. It is evident that, as shown in Fig.~\ref{fig5}, our present results have a remarkable overlap with the results obtained from Refs.~\cite{PhysRevC.83.054316,99xk-tm1q}.
Consequently, the constrained triplet-odd tensor strength parameter $U$ ranges from $-341$ MeV fm$^{5}$ to $-150$ MeV fm$^{5}$ for the effective SGII force. Meanwhile, for SLy5 force, the validated value of $U$ ranges from $-182$ MeV fm$^{5}$ to $-50$ MeV fm$^{5}$.
This means that the lower limit range of the triplet-odd tensor strength $U$ is further constrained.
Meanwhile, the constrained triplet-odd tensor strength $U$ can reproduce the absolute charge radii of $^{54}$Ni, $^{54}$Fe, $^{36}$Ca, and $^{36}$S well.
In Ref.~\cite{COLO2007227}, the constrained SLy5 tensor parameters have also been calibrated through the spin-orbit splitting in $Z=50$ isotopes and $N=82$ isotones. The calibrated value is overlapped with the triplet-odd tensor strength range obtained from the $\Delta{R_{ch}}$ of the mirror-paired nuclei $^{54}$Ni-$^{54}$Fe. While its value deviates from the $\Delta{R_{ch}}$ of the mirror-paired nuclei $^{36}$Ca-$^{36}$S. As is well known, tensor force can have an influence on determining the shell evolution. This tension should be further investigated by considering the bulk properties of finite nuclei, such as binding energies, charge radii, etc., in different shell regions.
Besides, as shown in Fig.~\ref{fig2}, the experimental uncertainty of $\pm0.004$ fm in mirror-paired nuclei $^{54}$Ni-$^{54}$Fe is relatively large to the spread of the theoretical lines, although the linear correlation can be presented between $\Delta{R}_{ch}$ and the triplet-odd tensor strengths $U$. This means that more mirror-paired nuclei data should be required.

Actually, the highly linear correlation between $\Delta{R}_{\mathrm{ch}}$ of mirror nuclei $^{48}$Ni-$^{48}$Ca and the triplet-odd tensor strength $U$ can also be found, but not be presented here. It should be remarkably mentioned that the triplet-even tensor strength $T$ cannot be constrained well due to the theoretical fitting lines are almost parallel to the x-axis with the increasing triplet-even tensor strength $T$. Furthermore, as encountered in $^{36}$Ca-$^{36}$S, the similar linear correlation can also be found in the mirror-paired nuclei $^{32}$Ar-$^{32}$Si. However, the linear correlation between $\Delta{R}_{\mathrm{ch}}$ of mirror-paired nuclei $^{32}$Ar-$^{32}$Si and the tensor strengths are distorted due to the absence of shape deformation effect. This means that more underlying mechanisms should be considered in constraining the tensor strength components. Meanwhile, more charge radii data of mirror-paired nuclei are urgently required in calibrating the tensor force. Particularly for mirror partner nuclei $^{48}$Ni-$^{48}$Ca, it seems to provide a potential access to calibrate the triplet-odd tensor strength $U$.

\section{Summary and Outlook}\label{sec4}
In this work, the correlation between the charge radii differences of mirror nuclei $^{54}$Ni-$^{54}$Fe and $^{36}$Ca-$^{36}$S and the tensor terms strength has been investigated with the effective forces SGII and SLy5 based on the Skyrme energy density functionals.
Highly linear correlations can be observed between the charge radii difference of mirror nuclei and the triplet-even tensor strength parameter $T$ as well as triplet-odd one $U$.
In contrast to the triplet-even tensor strength parameter $T$, triplet-odd tensor components are more sensitive to the values of $\Delta{R}_{\mathrm{ch}}$ of mirror nuclei.
Combining the existing results obtained from the charge-exchange Gamow-Teller (GT) and the spin-dipole (SD) excitations~\cite{PhysRevC.83.054316} and the magnetic dipole (M1) excitations~\cite{99xk-tm1q}, the triplet-odd tensor strength parameters $U$ are validated at the range from $-341$ MeV fm$^{5}$ to $-150$ MeV fm$^{5}$ for the effective SGII force and from the range of $-182$ MeV fm$^{5}$ to $-50$ MeV fm$^{5}$ for SLy5 set, respectively.
It should be mentioned that the calculated results have a remarkable overlap with the ranges shown in Refs.~\cite{PhysRevC.83.054316,99xk-tm1q} and the lower range is further constrained.
Meanwhile, it should be noted that the triplet-odd tensor strength $U$ depends on $T$ as well.

The induced tensor force can actually improve the description of bulk properties in finite nuclei~\cite{SUGIMOTO2004240}.
Meanwhile, the tensor force component can change the ground state deformation~\cite{PhysRevC.104.014313,BERNARD201632} or modify the effective interactions~\cite{JHA2020122038}.
As demonstrated in Ref.~\cite{PhysRevC.97.054325}, the already strong proton-neutron effective tensor interaction can be enhanced by the three-body forces, while the corresponding like-particle tensor force remains small.
Tensor correlation can shift nucleons to high momentum in symmetric nuclear matter due to the induced short-range correlations, but have almost no effect in pure neutron matter~\cite{PhysRevC.91.025803}.
Meanwhile, as demonstrated in Ref.~\cite{PhysRevC.81.044302}, adding tensor terms perturbatively to the existing effective forces lowers the critical density for spin-isospin instabilities in nuclear matter to values close to saturation density. The constrained tensor strengths may correspond to a regime where the functional is less reliable.
This means that a reliable tensor component play an indispensable role in characterizing the fundamental interactions in nuclear physics and astrophysics.
Furthermore, the systematic evolution of nuclear charge radii can be described well by considering the neutron and proton correlated pairs correction~\cite{nyn5-69s3}. It seems to be necessary to clarify the influence of neutron and proton correlation on determining the tensor components.
\section{Acknowledgements}\label{ackn}
This work was supported by the key research and development project of Ningxia, China (Nos. 2024BEH04090 and 2025BEH04008), the Natural Science Foundation of Ningxia Province, China (Nos. 2024AAC03015 and 2025AAC030208), the Central Government Guidance Funds for Local Scientific and Technological Development, China (No. Guike ZY22096024), the Open Project of Guangxi Key Laboratory of Nuclear Physics and Nuclear Technology (No. NLK2023-05), and the Key Laboratory of Beam Technology of Ministry of Education, China (Nos. BEAM2024G04 and BEAM2024G05).

\bibliography{refsanw}
\end{document}